\documentclass[aip,jmp,amsmath,amssymb,
 reprint,%
]{revtex4-1}

\bibpunct{}{}{,}{s}{}{}

\usepackage[latin1]{inputenc}
\usepackage[T1]{fontenc}

\usepackage{graphicx}
\usepackage{dcolumn}
\usepackage{bm}
\graphicspath{{fig/}}

\usepackage{units}
\usepackage{journames}

\usepackage[final]{neel}

\usepackage[mathlines]{lineno}
\pagewiselinenumbers 

\begin{document}

\renewcommand\topfraction{0.8}
\renewcommand\bottomfraction{0.7}
\renewcommand\floatpagefraction{0.7}

\preprint{AIP/123-QED}

\title{Epitaxial refractory-metal buffer layers with a chemical gradient for adjustable lattice parameter and controlled chemical interface}%

\author{O. Fruchart}
\email[]{Olivier.Fruchart@grenoble.cnrs.fr}
\author{A. Rousseau}
\affiliation{Institut N\'{e}el (CNRS and Universit\'{e} Joseph Fourier), BP166, F-38042 Grenoble Cedex 9, France}
\author{D. Schmaus}
\author{A. L'Hoir}
\affiliation{Institut des NanoSciences de Paris (INSP) (CNRS UMR 7588 and UPMC Universit\'{e} Paris 6), 4 place Jussieu, 75252 PARIS Cedex 05, France}
\affiliation{Universit\'{e} Paris Diderot-Paris 7, 75205 Paris Cedex 13, France}
\author{R. Haettel}
\author{L. Ortega}
\affiliation{Institut N\'{e}el (CNRS and Universit\'{e} Joseph Fourier), BP166, F-38042 Grenoble Cedex 9, France}

\date{\today}

\begin{abstract}

We have developed and characterized the structure and composition of nanometers-thick solid-solution epitaxial layers of (V,Nb) on sapphire ($11\overline{2}0$), displaying a
continuous lateral gradient of composition from one to another pure element. Further covered with an ultrathin pseudomorphic layer of W, these provide a template for the fast combinatorial investigation of any growth or physical property depending of strain.

\end{abstract}


\maketitle

\def\solsol{s.s.\/\xspace}
\def\CWL{CGL\/\xspace}
\def\CWLs{CGLs\/\xspace}

\vskip 0.5in

Thin films play a crucial role in integrated technology, and it is necessary to control their physical properties. In a down-scaling approach one seeks to sustain the bulk properties. One may also endeavor to tailor new properties that do not occur in the bulk, an approach being one basis for the development of nanosciences.
The parameter often at play in the change of properties in thin films is strain\cite{bib-NEW2005}, influencing \eg mobility in semiconductors\cite{bib-FIS1996}, optical activity, electric polarization or magnetic moment\cite{bib-MOR1989} and anisotropy\cite{bib-BUS2009}.
The easiest way to control strain in thin film is through their epitaxy on substrates with a lattice parameter different from their own, inducing a so-called lattice misfit. Lattice misfit also influences growth modes (smooth or rough films, islands...)\cite{bib-BAU1986,bib-MER1982}, which strengthens the need for its control.

Effects of lattice misfit are often investigated by growth on single crystals of various pure elements, each with a well-defined lattice parameter.
Alloys and solid solutions~(\solsol) are an appealing combinatorial alternative as they allow for the continuous control of lattice misfit, which is not possible for pure elements\cite{bib-FRI2000}.
Nevertheless, systematic studies with \solsol require several samples with different compositions, with issues of time, reproducibility and limited number of data points. To circumvent this Kennedy \etal\cite{bib-KEN1965} have introduced compositional spreads, where the anisotropy of evaporation of two sources are used to create a slight gradient of composition across a sample. Later, alternated deposition of the materials along with moving masks were used to create intercalated wedges of two or three elements\cite{bib-TSU2005,bib-MAT2007}. This allows one to vary the composition linearly and potentially from \unit[0]{\%} up to \unit[100]{\%}. In these seminal uses of masks the alloy resulted from a post-annealing, which is not suited for most epitaxial and thin films. Only
Zhong~\etal applied this technique to epitaxial materials (semiconductors), however restricted it to the case of weak doping\cite{bib-ZHO2007}.

We issued a preliminary report of chemical gradients with controlled chemical surface\cite{bib-FRU2007}, however with a negligible variation of lattice parameter via
mixing body-centered cubic~(bcc) Mo~(lattice parameter $\lengthAA{3.157}$) and W~($\lengthAA{3.165}$), therefore with virtually no  metallurgical issues and with
restricted practical use. Here we report the fabrication, the structural, composition and surface characterization of epitaxial Chemical-Gradient Layers (\CWLs) of
bcc refractory metals, in the full range of composition to adjust the lattice parameter over $\unit[10]{\%}$. Further combined with an ultrathin pseudomorphic layer, these provide a versatile buffer-layer toolkit for the elucidation and use of the many phenomena depending on lattice mismatch.

\begin{figure}
  \begin{center}
  \includegraphics[width=81mm]{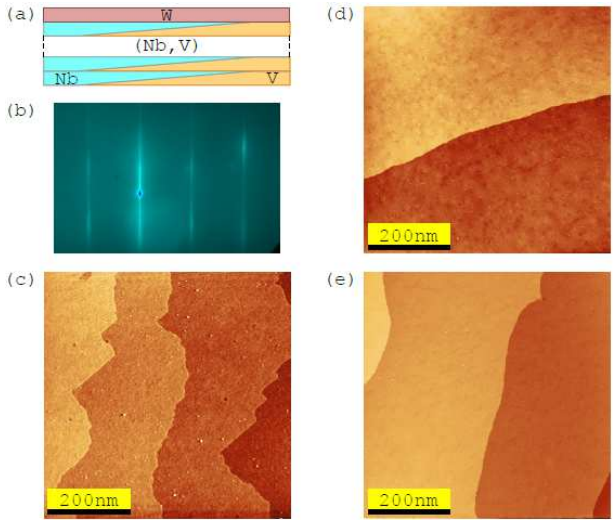}%
  \caption{\label{fig-rheed-stm}(a)~Schematics of a \CWL, further covered by an ultrathin W layer for the chemical control of the free surface. (b)~typical RHEED pattern, for electron azimuth $[1\overline{1}0]$ (the bright spot is the reflected beam). STM images of (V,Nb)/Mo(\lengthnm{0.7})/$\mathrm{Al}_2\mathrm{O}_3$ \CWLs (not covered with an ultrathin W layer) for (c)~V, (d)~$\mathrm{V}_{50}\mathrm{Nb}_{50}$ and
(e)~Nb . \dataref{Nb: AR17-313; Nb50V50: AR43-796; V: AR37-711. RHEED: AR17}}
  \end{center}
\end{figure}

The samples were grown using Pulsed-Laser Deposition in a set of Ultra-High-Vacuum chambers. The laser is a \unit[10]{Hz}-pulsed Nd-YAG laser with pulse length $\unit[\approx10]{ns}$ and doubled frequency~($\lambda=\lengthnm{532}$). The deposition chamber is equipped with a computer-controlled mask moving in front of the sample, and $\unit[10]{kV}$ Reflection High Energy Electron Diffraction~(RHEED)\cite{bib-FRU2007}. Our \CWLs are based on mixtures of two bcc elements~(Mo, W, Nb, V), that all form \solsol one with another. Results are illustrated in this Letter with the case of (V,Nb). $(11\overline{2}0)$ sapphire wafers are used as a support, resulting in the $(110)$ texture of the films\cite{bib-OYA1986,bib-FRU2007}. Searching for an optimized procedure, \CWLs were grown directly on $\lengthnm{0.7}$-thick-Mo-dusted sapphire wafer~(Mo is inserted to avoid crystallographic variants\cite{bib-FRU2007}), or on an atomically-flat $\lengthnm{10}$-thick W(110) buffer layer itself above Mo-dusted sapphire. The wedges of the two elements are deposited sequentially, in an opposite fashion thanks to the azimuthal rotation of the sample of $\angledeg{180}$. The length of the wedges is \unit[5]{mm}\bracketsubfigref{fig-rheed-stm}a. Their typical thickness is \thickAA1, which is less than one atomic layer, with a view to promote the mixture of both elements at the atomic level and avoid the formation of misfit dislocations that would occur for thick individual layers. Deposition is performed at moderate temperature~($\tempdegC{300}$), followed by annealing for \unit[30]{min} at $\tempdegC{800}$ to smoothen the surface.
The typical total thickness of these films is \lengthnm{10}.

\figref{fig-rheed-stm} shows RHEED patterns and Scanning Tunneling Microscopy~(STM) images of a (V,Nb)/Mo(\lengthnm{0.7}) \CWL directly grown on sapphire. The RHEED streak narrowness, large terrace size and absence of emerging screw dislocations, provide a picture of a high-quality crystal. No significant difference is found in RHEED and STM data as a function of the local composition, from pure V to pure Nb, with or without an underlying W buffer layer.

\begin{figure}
  \begin{center}
  \includegraphics[width=81mm]{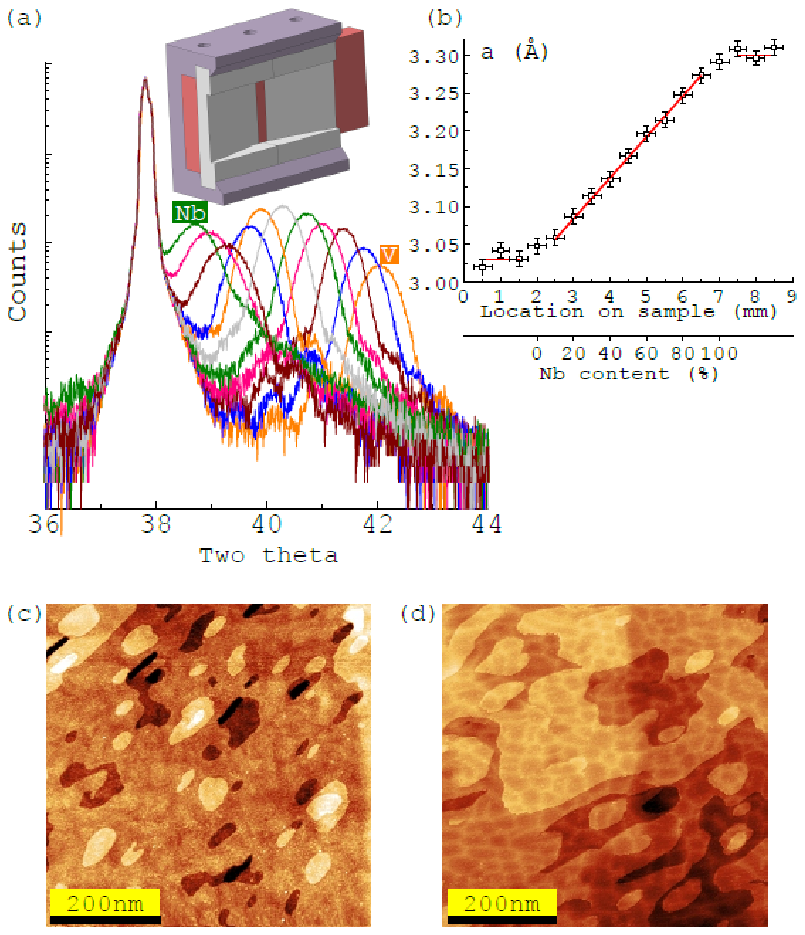}%
  \caption{\label{fig-rx}(a)~$\theta\mathrm{-}2\theta$ diffraction spectra of (V,Nb)/Mo(\lengthnm{0.7})/$\mathrm{Al}_2\mathrm{O}_3$ \CWLs. At
$\angledeg{37.85}$ is a sapphire peak\dataref{Spectres, FIG.p30 de rapport 5 Anthony}. Inset: mask used to select the area of interest for diffraction (b)~Extracted from (a), out-of-plane lattice parameter as a function of composition along the \CWL. Lines are the expectation of Vegard's law. STM images of ultrathin layers of W of thickness (c)~\lengthnm{0.6} and (d)~\lengthnm{2}
on a \CWL.\dataref{Couverture W: AR28, rapport 2bis, m590 pour 0.6nm, m598 pour 2nm}}
  \end{center}
\end{figure}

We developed a sample holder to perform X-Ray Diffraction over a small and precise location on the sample. It consists of a $\unit[1]{mm}$-wide window between two steel blades machined at an angle of $\angledeg{12}$ to let in and out X-rays down to grazing incidence. The blades are covered with a \lengthmicron{10}-thick layer of metallic glass~($\mathrm{Zr}_{52.5}\mathrm{Ti}_{2.5}\mathrm{Cu}_{22}\mathrm{Ni}_{13}\mathrm{Al}_{10}$) to avoid diffraction from the mask. A manual sample translator allows one to select the area of investigation~(inset of \subfigref{fig-rx}a).  \subfigref{fig-rx}a shows $\theta-2\theta$ spectra as a function of location on the sample, translated into the expected composition. The occurrence of Kiessig fringes arises from the finite thickness of the films and the absence of roughness. Their width is composition-independent and consistent with the film thickness. The films are thus coherent across their entire thickness, showing their good layered structure, consistent with STM data\bracketsubfigref{fig-rheed-stm}{c-e}. The out-of-plane lattice parameters extracted from these curves are shown in \figref{fig-rx}b. The error bars result from the precision of the lateral position of the sample, and from the combined fitting of peaks and calibration of the diffractometer against the first and second order peaks of sapphire. Within the error bar the lattice parameter varies linearly with composition and ranges from the bulk lattice parameter of V~(\lengthAA{3.02}) to that of Nb~(\lengthAA{3.30}), in agreement with Vegard's Law. \insitu RHEED yields the in-plane lattice parameter, albeit with larger error bars~(not shown here). These also fit Vegard's law, demonstrating the good structural relaxation of the \solsol. Finally, the \CWLs may be covered by an ultrathin W layer deposited at \tempdegC{250} and then annealed at \tempdegC{800}, a procedure known not to give rise to intermixing with the underlying film\cite{bib-FRU2007}. STM shows that W remains pseudomorphic up to $\lengthnm{\approx1}$\bracketsubfigref{fig-rx}{c}, above which relaxation is revealed by an array of misfit dislocations\bracketsubfigref{fig-rx}{d}\cite{bib-BET1995}. \CWLs covered with this ultrathin pseudomorphic W layer provide a surface with a continuously-variable lattice parameter, however with a uniform and a rather inert surface material.

In order to check locally both the composition of the \solsol, and their crystalline quality, the samples were also analyzed using Rutherford Backscattering Spectrometry (RBS) both in random and channeling geometries~(for these analytical techniques, see Ref.\cite{bib-SCH2006}). Indeed, whereas X-ray and electron diffraction probe long-range crystalline coherence and are mostly insensitive to local disorder, RBS in a channeling geometry is highly sensitive to such disorder and is thus very valuable to refine the close-to-perfect picture of \CWLs given by \figref{fig-rheed-stm} and \ref{fig-rx}. RBS was performed within the SAFIR facility of the INSP, using a \unit[1.4]{MeV} $^4\mathrm{He}^{+}$ ion beam with a $\unit[0.5]{mm}$ diameter, produced by a \unit[2.5]{MV} Van de Graaff accelerator. The ions scattered elastically at large angle (here $\approx\angledeg{150}$) on the sample nuclei (rare scattering events) were energy-analyzed using a silicon detector.

Conversely, the most probable scattering events result in small-angle repulsive deflections. When the incident beam is aligned with a major crystallographic axis of a crystal (channeling geometry, here performed along $[110]$), these deflections are correlated and the particle flux close to the rows is strongly reduced, leading to a similar reduction of the RBS yield $Y$ with respect to the yield $R$ using a random orientation of the beam. For convenience, in the following $Y$ values are normalized to the corresponding RBS random yield. Whereas $R_i$ values for element $i$ are proportional to the absolute amount of $i$ atoms in the layer, $Y_i$ values are mostly sensitive to atomic displacements in the plane perpendicular to the channeling axis\cite{bib-rbs-detailed}.

\begin{figure}
  \begin{center}
  \includegraphics[width=70.759mm]{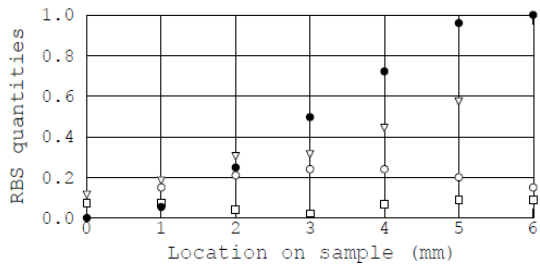}%
  \caption{\label{fig-rbs}Local quantities determined by RBS~(random or channeling conditions), as a function of the location $x$ across a (V,Nb)/W(\lengthnm{10})/Mo(\lengthnm{0.7})/$\mathrm{Al}_2\mathrm{O}_3$ \CWL : Nb concentration~($\bullet$), normalized channeling ($[110]$ alignment) yields $Y_\mathrm{Nb}$~($\circ$), $Y_\mathrm{V}$~($\triangledown$) and $Y_\mathrm{W}$~($\scriptstyle\square$)~(see text). The zero for the lateral scale is different from that of \subfigref{fig-rx}b.}
  \end{center}
\end{figure}

Various samples were analyzed, but we only present here the case of (V,Nb) deposited on a $\lengthnm{10}$-thick W(110) buffer layer\bracketfigref{fig-rbs}. Both $R_\mathrm{V}(x)$ and $R_\mathrm{Nb}(x)$ exhibit a linear variation across the sample, their sum keeping a nearly constant value corresponding to a thickness of about $\lengthnm{10}$ of the (Nb,V) layer. This provides a direct quantitative confirmation of the linear variation of composition $[$\eg $C_\mathrm{Nb}=R_\mathrm{Nb}/(R_\mathrm{Nb}+R_\mathrm{V})]$ with $x$, suggested indirectly previously through the agreement with Vegard's Law\bracketfigref{fig-rx}.

Concerning the crystalline quality of the \solsol, the relatively high channeling yields $Y_\mathrm{Nb}$ and $Y_\mathrm{V}$ observed on \figref{fig-rbs}, with a marked dependence on $C_\mathrm{Nb}$, may hint at atomic displacements in the \solsol, relatively far from the \CWL-W interface~(see below). Their precise identification needs further analysis. Despite this, paradoxically, the very low channeling yield $Y_\mathrm W$ in the range $[0.02\mathrm{-}0.05]$ is indicative of a very good epitaxy. The highest quality~($Y_\mathrm W\approx0.02$) is obtained for $\mathrm{Nb}_{0.5}\mathrm{V}_{0.5}$, for which there is no lattice mismatch between the alloy and W, in agreement with \subfigref{fig-rx}b.

To conclude, we developed high-quality epitaxial solid-solution layers of refractory metals along the orientation (110), in the form of Chemical-Gradient Layers (\CWLs). These \CWLs were then covered with a flat and ultrathin pseudomorphic layer of W(110). This combination provides a template with a lattice parameter varying laterally over $\unit[10]{\%}$ and with a controlled and rather inert W chemical interface, which may be used for the fast combinatorial investigation of any growth or physical phenomenon depending on strain. This may be performed with local probes such as electron, optical or scanning-probe microscopies, or with many nowadays synchrotrons offering beams focused to \lengthmicron{100} or smaller.

\vskip 0.5in


\section*{Acknowledgments}

We acknowledge the contribution of J. L. Soubeyroux~(CRETA, Grenoble) for providing bulk metallic glass for
preliminary XRD slits, N. Dempsey~(Institut N\'{e}el) for the deposition of thick films of metallic glass, V.~Guisset and Ph.~David for technical support with UHV and E.~Briand for efficient help in RBS experiments. This work received financial support from FP6 EU-NSF program (STRP 016447 MagDot) and French National
Research Agency (ANR-05-NANO-073 Vernanomag).

\section*{References}


\begin{thebibliography}{16}
\bibitem [{bib()}]{bib-rbs-detailed}%
  \BibitemOpen
  \href@noop {} {}\bibinfo {note} {A detailed discussion of RBS-channeling data
  will be addressed separately}\BibitemShut {NoStop}%
\end{thebibliography}

%

\end{document}